\documentclass[12pt]{article}
\usepackage[latin9]{inputenc}
\usepackage{amsmath}
\usepackage{amssymb}
\usepackage{graphicx}
\usepackage{placeins}
\usepackage{lineno}
\usepackage{titlesec}

\titleformat{\subsection}[runin]{\bfseries}{}{0.0em}{}
\newcommand{\beq}{\begin{equation}}
\newcommand{\eeq}{\end{equation}}
\newcommand{\beqa}{\begin{eqnarray}}
\newcommand{\eeqa}{\end{eqnarray}}

\begin{document}

\title{Localized solutions of Lugiato-Lefever equations with focused pump}
\author{Wesley B. Cardoso$^{1}$, Luca Salasnich$^{2,3,*}$, Boris A. Malomed$%
^{4,5}$}
\maketitle
\noindent
$^{1}$Instituto de F\'{i}sica, Universidade Federal de Goi\'{a}s, 74.690-900,
Goi\'{a}nia, Goi\'{a}s, Brazil \\[0pt]
$^{2}$Dipartimento di Fisica e Astronomia ``Galileo
Galilei\textquotedblright \ and CNISM, Universit\'{a} di Padova, Via Marzolo 8,
35131 Padova, Italy \\[0pt]
$^{3}$Istituto Nazionale di Ottica del Consiglio Nazionale delle Ricerche,
Via Nello Carrara 1, 50019 Sesto Fiorentino, Italy \\[0pt]
$^{4}$Department of Physical Electronics, School of Electrical
Engineering, Faculty of Engineering, Tel Aviv University, Tel Aviv 69978,
Israel\\[0pt]
$^{5}$ITMO University, St. Petersburg 197101, Russia \\[0pt]
$^{*}$Correspondence to luca.salasnich@unipd.it

\begin{abstract}
Lugiato-Lefever (LL) equations in one and two dimensions (1D and 2D)
accurately describe the dynamics of optical fields in pumped lossy cavities
with the intrinsic Kerr nonlinearity. The external pump is usually assumed
to be uniform, but it can be made tightly focused too --{} in particular,
for building small pixels. We obtain solutions of the LL equations, with
both the focusing and defocusing intrinsic nonlinearity, for 1D and 2D
confined modes supported by the localized pump. In the 1D setting, we first
develop a simple perturbation theory, based in the \textrm{sech} \textit{%
ansatz}, in the case of weak pump and loss. Then, a family of exact
analytical solutions for spatially confined modes is produced for the pump
focused in the form of a delta-function, with a nonlinear loss (two-photon
absorption) added to the LL model. Numerical findings demonstrate that these
exact solutions are stable, both dynamically and structurally (the latter
means that stable numerical solutions close to the exact ones are found when
a specific condition, necessary for the existence of the analytical
solution, does not hold). In 2D, vast families of stable confined modes are
produced by means of a variational approximation and full numerical
simulations. 
\end{abstract}

It is commonly known that stable self-confined modes, such as solitons, may
be produced by the balance between nonlinear and dispersive effects in the
medium \cite{Zabusky_PRL65}. Solitons have been observed in diverse
contexts, including water waves \cite{Craig_PF06}, nonlinear fiber optics
\cite{Hasegawa_95,Agrawal_13} (as temporal solitons), Bose-Einstein
condensates (BECs) \cite{Khaykovich_SCI02,Strecker_NAT02,Heidelberg,
Cornish_PRL06,Marchant_NC13,Burger_PRL99,Romania}, plasmas \cite%
{plasma1,plasma2} and plasmonics \cite{plasmonic1,plasmonic2}, proteins \cite%
{Davydov_85} and DNA \cite{Yakushevich_04}, \textit{etc}. Optical spatial
solitons were created too in a great variety of settings, such as cells
filled by vapors of alkali metals \cite{Bjorkholm_PRL74}, photorefractive
crystals \cite{Segev_PRL92,Segev-OE}, waveguides made of liquid dielectrics
\cite{Barthelemy_OC85,Cid}, silica \cite{Silberberg} and
second-harmonic-generating materials \cite{chi2}, nematic liquid-crystal
planar cells \cite{Beeckman_OE04}, semiconductor waveguides \cite%
{Aitchison_EL92}, arrayed waveguides \cite{array}, and others.

While solitons have been originally introduced as exact solutions of
integrable models \cite{Zabusky_PRL65,Zakharov,Sulem_07}, nonintegrable
systems provide for more generic and more realistic description of various
physical settings. In particular, numerous dissipative systems, although
lacking integrability, readily give rise to robust localized dissipative
structures (LDSs), alias dissipative solitons \cite%
{Akhmediev_08,Purwins_AP10}. In optics, an important example of a nonlinear
dissipative medium which supports LDSs is provided by an optical resonator
filled with a dispersive loss material featuring the Kerr nonlinearity,
which is pumped by a coherent light beam (injected signal). This system is
well modeled by the Lugiato-Lefever (LL) equation, originally introduced in
\cite{Lugiato_PRL87}. As a mean-field equation, it applies to other settings
too, such as Fabry-Perot resonators and ring cavities, fully or partially
filled with nonlinear materials \cite{Lugiato_PRL87}, crystalline
whispering-gallery-mode disk resonators \cite{Coillet_IEEE13}, and
photonic-crystal-fiber resonators pumped by a coherent continuous-wave input
beam \cite{Tlidi_OL07,Tlidi_OL10}. In these contexts, the LL equation has
been widely used to model Kerr frequency combs \cite%
{Coen_OL13,Chembo_PRA13,Pfeifle_PRL15,Loures_PRL15,Hansson_NanoP16}, with
applications to optical metrology \cite{Cundiff_RSI01}, high-precision
spectroscopy \cite{Kippenberg_SCI11,Cingoz_NAT12}, optical atomic clocks
\cite{Diddams_SCI01,Ye_PRL01}, phase evolution in pulse trains \cite%
{Apolonski_PRL00,Jones_SCI00}, optical communications \cite{Temprana_SCI15},
synthesis of arbitrary optical waveforms \cite{Shelton_SCI01,Cundiff_NP10},
and radio-frequency photonics \cite{Torres-Company_LPR14}. A review of the
development of various applications of the LL equations has been published
recently \cite{LL-history}. Soliton-like LDSs in the 1D LL equation are
important modes too, in a broad range of values of the respective physical
parameters \cite{LLsolitons}.

In many physically relevant contexts, especially as concerns realizations in
optics, one- and two-dimensional (1D and 2D) LDSs supported by localized
gain were studied in detail in the context of complex Ginzburg-Landau (CGL)
equations \cite{Lam_EPJST09,Kartashov_OL10,Kartashov_OL11,
Lobanov_OL11,Zezyulin_OL11,Borovkova_OL11,Ye_OL13,Huang_OL13,Lobanov_PRA14},
see also reviews in \cite{in-book} and \cite{Malomed_JOSAB14}. Specifically,
by considering a 1D model with the tightly localized gain in the form of a
delta-function, placed on top of the spatially uniform linear loss,
analytical solutions for pinned LDSs pinned to the delta-function were found
in \cite{Lam_EPJST09,Ye_OL13} (see also a review in \cite{Malomed_JOSAB14}).
Stable LDSs pinned to one or two gain-carrying \textquotedblleft hot spots",
shaped as narrow Gaussians, were reported too \cite%
{Zezyulin_OL11,Lobanov_PRA14}. Further, stable 2D LDSs, including ones with
an intrinsic vortex structure, supported by hot spots in the 2D geometry,
were predicted in works \cite%
{Kartashov_OL10,Kartashov_OL11,Lobanov_OL11,Borovkova_OL11,Huang_OL13}.

The objective of the present work is to introduce localized pump in the
framework of the 1D and 2D LL equations, and find stable confined modes,
which may be supported by the spatially focused pump. The difference from
the previous works, which were dealing with the CGL equations \cite%
{Lam_EPJST09,Kartashov_OL10,Kartashov_OL11,
Lobanov_OL11,Zezyulin_OL11,Borovkova_OL11,Ye_OL13,Huang_OL13,Malomed_JOSAB14, Lobanov_PRA14}%
, is that the pump is represented by free terms in the LL equations, which
do not multiply the field variable, while in the models of the CGL type the
gain \ terms provide the parametric pump, i.e., they multiply the field
variable.

We report analytical solutions of the 1D and 2D LL equation with the
localized external pump, using a possibility to find exact analytical
solutions for 1D modes pinned to the pump represented by the delta-function,
and a variational approach, respectively. In the case of weak pump and loss,
a simple perturbation theory for 1D modes is developed too. In a systematic
form, the results for confined modes, including the analysis of their
stability, are produced by means of numerical methods. The results
demonstrate good agreement between the analytical predictions and numerical
findings. In particular, while the exact analytical solutions for pinned
modes in 1D are available under a special condition, we demonstrate that
very similar stable numerical solutions exist when this condition does not
hold. The predicted confined stable modes, pinned to the \textquotedblleft
hot spots", may be used, in particular, for the design of pixels placed at
required positions, cf. the formation of pixels predicted by the LL equation
in other contexts \cite{pixel}.

\section*{Results}

\subsection*{The one-dimensional Lugiato-Lefever equation.}

\textbf{The 1D model equations}. In the 1D\ setting, the scaled LL equation
for amplitude $\phi (x,t)$ of the electromagnetic field in a nonlinear lossy
cavity driven by a real localized pump $E(x)$ is
\begin{equation}
i\left( \gamma +\frac{\partial }{\partial t}\right) \phi =\left[ -\frac{1}{2}%
\frac{\partial ^{2}}{\partial x^{2}}+\Delta +\sigma |\phi |^{2}\right] \phi
+E(x),  \label{LuLe-1D}
\end{equation}%
where $\gamma >0$ is the dissipation rate, $\Delta $ is detuning of the pump
with respect to the cavity, while $\sigma =-1$ and $+1$ corresponds to the
self-focusing and defocusing nonlinearity, respectively. Note that
dissipative solitons in the model of a fiber cavity, based on the 1D\ LL
equation written in the temporal domain, with a pump in the form of a period
train of Gaussians pulses, placed on top a nonzero background, were recently
considered in work \cite{Parra-Rivas_OE14}. Accordingly, the LL equation (%
\ref{LuLe-1D})\ may also be considered in the temporal domain, with $t$ and $%
x$ replaced, respectively, by the propagation distance ($z$) and the
temporal coordinate (usually denoted $\tau $).

Stability of various patterns produced by Eq. (\ref{LuLe-1D}) and its 2D
counterpart considered below may be enhanced if an extra cubic lossy term,
which represents the two-photon absorption, is added to the model. Then, Eq.
(\ref{LuLe-1D}) is replaced by
\begin{equation}
i\left(\gamma+\Gamma|\phi|^{2}+\frac{\partial}{\partial t}\right)\phi=\left[-%
\frac{1}{2}\frac{\partial^{2}}{\partial x^{2}}+\Delta+\sigma|\phi|^{2}\right]%
\phi+E(x),  \label{New_LuLe-1D}
\end{equation}
with $\gamma,\Gamma>0$.

Before proceeding to analysis of confined modes supported by the tightly
localized pump, it is relevant to mention that spatial localization may also
be provided, in the presence of the usual uniform pump, by a confining
(typically, harmonic-oscillator) potential \cite{we}. On the other hand,
results reported below demonstrate that the use of the narrow pump region
does not imply that modes supported by it must necessarily be narrow too.
Note that effects of local defects on LDSs in similar settings were
previously studied in works \cite{defect1} and \cite{defect2}.

\textbf{The perturbative treatment.} In the case of the self-focusing
nonlinearity ($\sigma =-1$) and positive detuning, $\Delta >0$, one can
develop a perturbation theory for the case of small $\gamma $ and small $E(x)
$ in Eqs. (\ref{LuLe-1D}) and (\ref{New_LuLe-1D}). In the zero-order
approximation, a localized solution is given by the usual nonlinear-Schrö%
dinger soliton \cite{Zakharov},
\begin{equation}
\phi (x)=e^{-i\zeta }\ \sqrt{2\Delta }\ \mathrm{sech}\left( \sqrt{2\Delta }%
x\right)   \label{soliton}
\end{equation}%
(as the zero-order approximation for localized patterns in the LL equation
with $E=\mathrm{const}$, the soliton waveform was used before \cite%
{Jang_NJP16}). The constant phase shift in \textit{ansatz} (\ref{soliton}), $%
\zeta $, for stationary modes is then determined by the balance condition
for the integral power,
\begin{equation}
P=\int_{-\infty }^{+\infty }\left\vert \phi (x)\right\vert ^{2}dx.  \label{P}
\end{equation}%
Indeed, it follows from condition $dP/dt=0$ that
\begin{equation}
\gamma P+\Gamma \int_{-\infty }^{+\infty }\left\vert \phi (x)\right\vert
^{4}dx=-\int_{-\infty }^{+\infty }E(x)\mathrm{Im}\left\{ \phi (x)\right\} dx.
\label{balance}
\end{equation}%
Substituting the \textrm{sech} approximation (\ref{soliton}) into Eq. (\ref%
{balance}) predicts the value of the phase shift:
\begin{equation}
\sin \zeta =\frac{2\left[ \gamma +\left( 4/3\right) \Gamma \Delta \right] }{%
\int_{-\infty }^{+\infty }E(x)\ \mathrm{sech}\left( \sqrt{2\Delta }x\right)
dx}~,  \label{sin}
\end{equation}%
which is written for the generalized LL equation (\ref{New_LuLe-1D}), that
includes the cubic loss $\sim \Gamma $. This result makes sense if it yields
$\left\vert \sin \zeta \right\vert \leq 1$, which implies that the LDS of
the prsent type exists if the pump's strength exceeds a threshold value,
which is a combination of dissipation coefficients $\gamma $ and $\Gamma $.
In fact, a mode pinned to the localized pump exists at all values of its
strength, as demonstrated by the exact solution displayed below, the
threshold being an artifact following from the assumption of the rigid form
of the perturbative ansatz (\ref{soliton}).

Note that, even for $E(x)=\mathrm{const}\equiv \mathcal{E}_{0}$, integral $%
\int_{-\infty }^{+\infty }E(x)\mathrm{sech}\left( \sqrt{2\Delta }x\right) dx$
converges , hence the approximation based on Eqs. (\ref{soliton})-(\ref{sin}%
) may correctly predict a state sitting on top of a small-amplitude CW
background, with amplitude $\phi _{0}\approx \mathcal{E}_{0}/\left( \Delta
+i\gamma \right) $, under the condition that the LDS's amplitude, $\sqrt{%
2\Delta }$, is much larger than $\phi _{0}$, i.e., $\mathcal{E}_{0}^{2}\ll
\Delta ^{3}$. Detailed comparison of predictions of the perturbation theory
with numerical results will be presented elsewhere.

\textbf{A particular exact solution and states close to it.} In the case
when the gain is localized in a very narrow region, it may be approximated
by the Dirac's delta-function, cf. a similar approximation adopted for a
strongly localized gain in the CGL model \cite{Lam_EPJST09}:
\begin{equation}
E(x)=E_{0}\ \delta (x)  \label{(x)}
\end{equation}%
(a similar model including a localized gain, with an LDS pinned to it, was
also formulated in terms of the Swift-Hohenberg equation \cite{SH}). This
means that the homogeneous version of Eq. (\ref{New_LuLe-1D}),
\begin{equation}
i\left( \gamma +\Gamma |\phi |^{2}+\frac{\partial }{\partial t}\right) \phi =%
\left[ -\frac{1}{2}\frac{\partial ^{2}}{\partial x^{2}}+\Delta +\sigma |\phi
|^{2}\right] \phi ,  \label{homo}
\end{equation}%
must be solved with the boundary condition at $x=0$ which determines the
jump of the first derivative induced by $\delta (x)$ in Eq. (\ref{(x)}):
\begin{equation}
\frac{d\phi }{dx}|_{x=+0}-\frac{d\phi }{dx}|_{x=-0}=2E_{0}.  \label{jump}
\end{equation}%
In this case, one can find a particular exact solution to the generalized LL
equation (\ref{homo}) in the form of
\begin{equation}
\phi (x)=\frac{Ae^{i\zeta }}{\left[ \sinh \left( \lambda \left( |x|+\xi
\right) \right) \right] ^{1+i\mu }},  \label{sinh}
\end{equation}%
with parameters ($\mu $ is called the \textit{chirp})
\begin{gather}
\mu =-\gamma /\lambda ^{2},  \label{mu} \\
A^{2}=3\gamma /\left( 2\Gamma \right) ,  \label{A^2} \\
\lambda ^{2}=\frac{\gamma }{4\Gamma }\left( \sqrt{9\sigma ^{2}+8\Gamma ^{2}}%
+3\sigma \right) .  \label{lambda^2}
\end{gather}%
This particular solution is a non-generic one, as it exists at the \textit{%
single value} of the mismatch parameter,
\begin{equation}
\Delta =\frac{\lambda ^{2}}{2}\left( 1-\mu ^{2}\right) \equiv \frac{1}{2}%
\left( \frac{3\sigma \gamma }{\Gamma }-\lambda ^{2}\right)   \label{mism}
\end{equation}%
(in other words, it is a\textit{\ codimension-one} type of the exact
solution, with \textquotedblleft one\textquotedblright\ referring to
constraint (\ref{mism}), which must be adopted to produce the analytical
expression). Note that the solution given by Eq. (\ref{lambda^2}) exists
(i.e., it gives $\lambda ^{2}>0$) for both $\sigma =-1$ and $+1$. The
presence of the cubic-loss coefficient, $\Gamma >0$, is necessary for the
existence of the solution. Indeed, in the limit of $\Gamma \rightarrow 0$
Eq. (\ref{lambda^2}) leads to divergence:
\begin{equation}
\lambda ^{2}\approx \left\{
\begin{array}{c}
3\gamma /\left( 2\Gamma \right) ~~\mathrm{at}~~\sigma =+1, \\
(1/3)\Gamma \gamma ~~\mathrm{at}~~\sigma =-1.%
\end{array}%
\right.   \label{to 0}
\end{equation}

Finally, parameters $\xi $ and $\zeta $ in expression (\ref{sinh}) are
obtained by its substitution in jump condition (\ref{jump}):
\begin{equation}
A\lambda \left( 1+i\mu \right) e^{i\zeta \lambda }\frac{\cosh \left( \lambda
\xi \right) }{\left[ \sinh \left( \lambda \xi \right) \right] ^{2+i\mu }}%
=-E_{0}.  \label{E0}
\end{equation}%
An explicit result, following from Eq. (\ref{E0}), is
\begin{equation}
\xi =\frac{1}{2\lambda }\mathrm{arcosh}\left( 1+\frac{\chi }{E_{0}^{2}}+%
\sqrt{4+\frac{\chi }{\lambda ^{2}E_{0}^{4}}}\right) ,  \label{xi}
\end{equation}%
\begin{equation*}
\chi \equiv A^{2}\lambda ^{2}\left( 1+\mu ^{2}\right) ,
\end{equation*}%
\begin{equation}
\zeta =\pi -\arctan \mu +\mu \ln \left( \sinh \left( \lambda \xi \right)
\right) ,  \label{d}
\end{equation}%
where $\mathrm{arcosh}(Z)~\equiv \ln \left( Z+\sqrt{Z^{2}-1}\right) $.

In the CGL\ model with localized gain (rather than pump), exact pinned
states are also codimension-one solutions, the difference being that, in the
latter case they coexist with the zero state, which may or may not be stable
solutions \cite{Lam_EPJST09}.

\begin{figure}[tb]
\centering \includegraphics[width=1.\columnwidth]{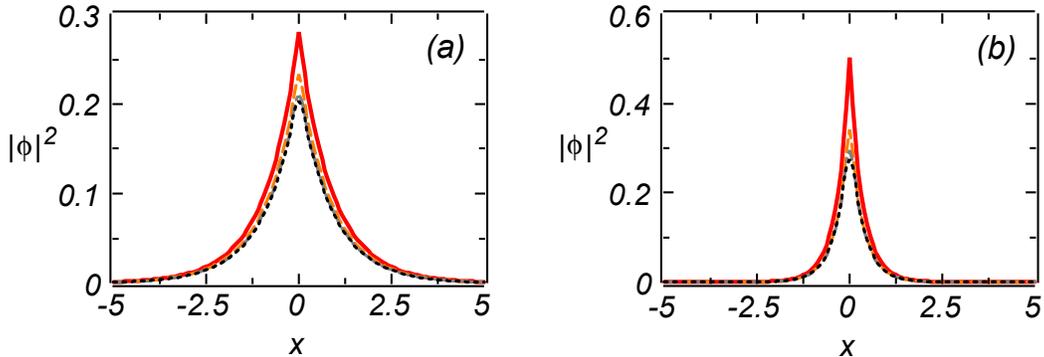}
\caption{Solid red lines display the exact solution (\protect\ref{sinh}) for
the mode pinned to the delta-functional pump, and a set of numerical
solutions based on the use of the regularized delta-function defined as per
Eq. (\protect\ref{delta}) (see \textbf{Methods}), with $w=0.05$ (dashed
orange lines), $w=0.1$ (dashed-dotted gray lines), and $w=0.15$ (dotted
black lines). The results presented in (a) and (b) pertain to self-focusing (%
$\protect\sigma =-1$) and self-defocusing ($\protect\sigma =1$) signs of the
nonlinearity, respectively. All these solutions are stable. Other parameters
are $E_{0}=\protect\gamma =\Gamma =1$, while $\Delta $ is given by Eq. (%
\protect\ref{mism}). }
\label{1DF1}
\end{figure}

In Fig. \ref{1DF1}(a)-(b) we display typical examples of the analytically
found modes pinned to the delta-function for focusing and defocusing
nonlinearities, by choosing $\sigma =-1$ and $\sigma =1$, respectively,
along with their numerically found counterparts. In this case, we set
parameters as $\gamma =\Gamma =E_{0}=1$, and took $\Delta $ as per Eq. (\ref%
{mism}). The numerical counterparts were produced by using the naturally
regularized delta function in the form given by Eq. (\ref{delta}) (see
\textbf{Methods}), for three different values of width $w$. It is relevant
to mention that, while the regularized delta-function approaches the
standard delta-function in the limit of $w\rightarrow 0$, the use of a
finite stepsize $\Delta x$ in the numerical procedure gives rise to a
critical value $w_{\mathrm{cr}}\simeq \Delta x/2$ of $w$, the numerical
solution getting drastically different from the analytical one at $w<w_{%
\mathrm{cr}}$. With the increase of the cubic-loss strength, $\Gamma $, $w_{%
\mathrm{cr}}$ blows up (increases very fast) at $\Gamma \gtrsim 3$.

\begin{figure}[tb]
\centering \includegraphics[width=1.\columnwidth]{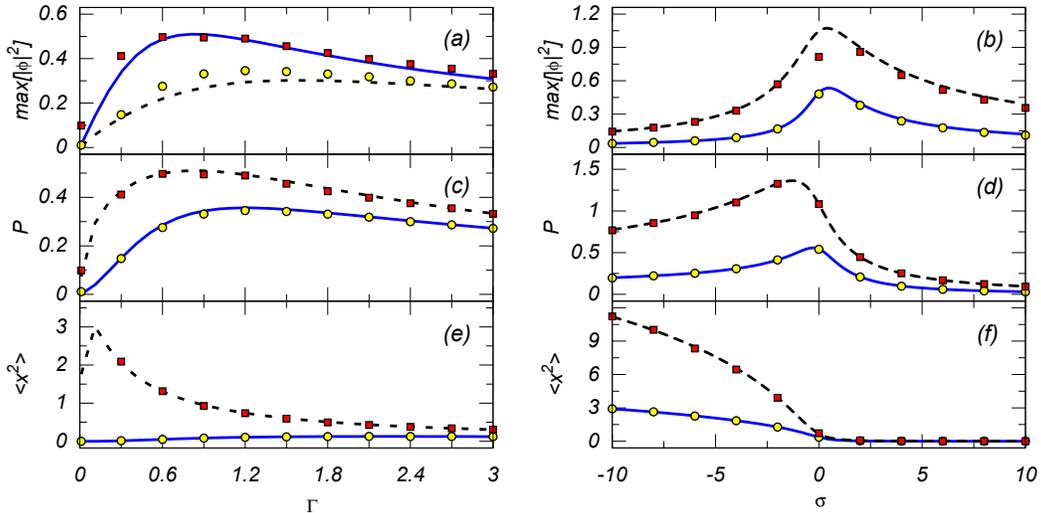}
\caption{Panels (a,b), (c,d), and (e,f) show, severally, the peak local
power, $\mathrm{max}[|\protect\phi |^{2}]$, integral norm $P$ (see Eq. (%
\protect\ref{P})), and the mean squared width $\langle x^{2}\rangle $ (see
Eq. (\protect\ref{<>})) of the analytical mode (\protect\ref{sinh}) versus
the cubic-loss and self-interaction strengths, $\Gamma $ and $\protect\sigma
$ (left and right columns). The left columns correspond to $\protect\gamma %
=E_{0}=1$ and $\protect\sigma =1$ (solid blue lines) or $\protect\sigma =-1$
(dashed black lines), i.e., the self-defocusing and focusing, respectively.
In the right columns we set $\protect\gamma =\Gamma =1$ and $E_{0}=1$ (solid
blue lines) or $E_{0}=2$ (dashed black lines). The corresponding numerical
results are shown by chains of yellow circles and red boxes, respectively.
The numerical data displayed here and other figures have been produced using
the regularized delta-function (\protect\ref{delta}), with $w$ close to its
above-mentioned critical value (see \textbf{Methods}).}
\label{1DF2}
\end{figure}

Further, in Fig. \ref{1DF2} we present systematic results for the 1D modes
produced by analytical solution (\ref{sinh}) and its numerical counterparts.
These are the peak local power, $\mathrm{max}\left[ |\phi |^{2}\right] $,
the integral power, $P$ (see Eq. (\ref{P})), and the mean squared width,
\begin{equation}
\langle x^{2}\rangle =P^{-1}\int_{-\infty }^{+\infty }x^{2}\left\vert \phi
(x)\right\vert ^{2}dx,  \label{<>}
\end{equation}%
shown in the left and right panels of the figure, as functions of the two
nonlinearity coefficients, \textit{viz}., the cubic-loss strength $\Gamma $
and self-interaction strength $\sigma $ (in the right panel, $\sigma $ is
considered as a continuously varying parameter, while in the left panel it
is fixed to be $\sigma =\pm 1$ for the self-defocusing and focusing cases).
Note that, as predicted by the analytical solutions (see Eqs. (\ref{sinh})
and (\ref{to 0})), the integral power $P$ vanishes at $\Gamma \rightarrow 0$%
. On the other hand, the results pertaining to $\sigma =+1$ and $-1$ tend to
converge at large values of $\Gamma $, as the dissipative nonlinearity is
dominant in this limit. The numerically generated findings are very close to
the analytical predictions.

It is worthy to note conspicuous maxima of the peak local power and integral
power, observed in Fig. \ref{1DF2}(b) at $\sigma=0$ and $\sigma\approx-1$,
respectively. Further, the bottom panel in Fig. \ref{1DF2}(b) reveals a
\textit{counter-intuitive} feature of the pinned states: they shrink ($%
\langle x^{2}\rangle\rightarrow0$) in the limit of large $\sigma>0$, i.e.,
\textit{strong self-defocusing} (the same is also demonstrated by Eq. (\ref%
{lambda^2}), which predicts $\lambda^{2}\sim1/\langle
x^{2}\rangle\rightarrow\infty$ at $\sigma\rightarrow+\infty$). Usually,
self-confined modes shrink in the opposite limit, of strong self-focusing.
This surprising finding may be explained by the effect introduced by the
cubic loss term $\sim\Gamma$. Indeed, as mentioned above, the exact solution
for the pinned state does not exist at $\Gamma=0$, and, in the presence of $%
\Gamma>0$, the shape of the mode is essentially affected by its chirp, which
is produced by Eq. (\ref{mu}).

\begin{figure}[tb]
\centering
\includegraphics[width=1.\columnwidth]{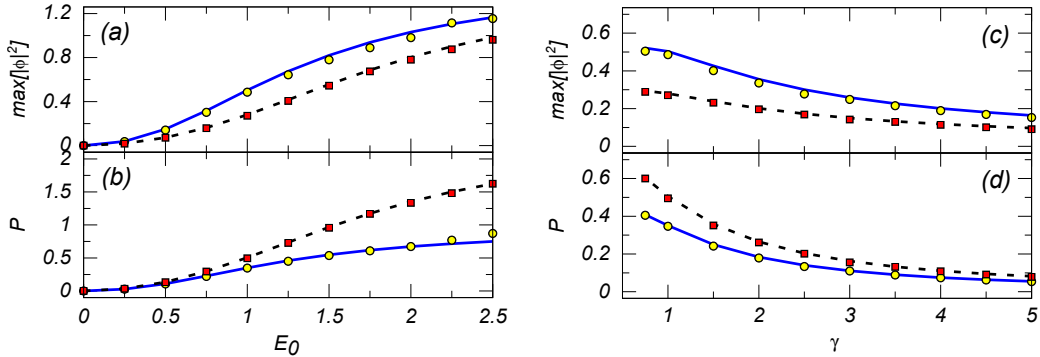}
\caption{Panels (a,c) and (b,d) show the peak local power ($\mathrm{max}[|%
\protect\phi |^{2}]$) and integral norm ($P$, see Eq. (\protect\ref{P})),
respectively, versus the pumping amplitude $E_{0}$ and the dissipation rate $%
\protect\gamma $ (left and right columns). The left columns correspond to $%
\protect\gamma =\Gamma =1$ while in the right columns we set $E_{0}=\Gamma
=1 $, both with $\protect\sigma =+1$ and $\protect\sigma =-1$, i.e., the
self-defocusing and focusing (the corresponding analytical results are
displayed by solid blue lines and dashed black lines, respectively), while
the corresponding numerical results are shown by chains of yellow circles
and red boxes, respectively.}
\label{NE3}
\end{figure}

Further, Fig. \ref{NE3} displays the effect of the variation of the pump's
amplitude $E_{0}$ and dissipation coefficient $\gamma $ on the peak local
power (max[$|\phi|^{2}$]) and integral power $P$ (see Eq. (\ref{P})) of
numerical solutions obtained from Eq. (\ref{New_LuLe-1D}), for both the
self-defocusing and focusing signs of the nonlinearity, i.e., $\sigma =+1$
and $\sigma =-1$, respectively, along with the counterparts predicted by the
above analytical solutions. Naturally, the peak local and integral powers
increase with the growth of $E_{0}$, and decrease with the growth of $\gamma
$. These properties can be used for an effective control of the localized
modes by means of parameters $E_{0}$ and $\gamma $.

Comparing the results obtained for the self-defocusing ($\sigma =+1$) and
self-focusing ($\sigma =-1$) signs of the nonlinearity, we again observe a
\textquotedblleft counter-intuitive" phenomenon, similar to that mentioned
above, i.e., the solution is more localized in the case of the
self-defocusing case than in the self-focusing case. Note that numerical
results closely follow their analytical counterparts in Fig. \ref{NE3} too.

\begin{figure}[tb]
\centering
\includegraphics[width=1.\columnwidth]{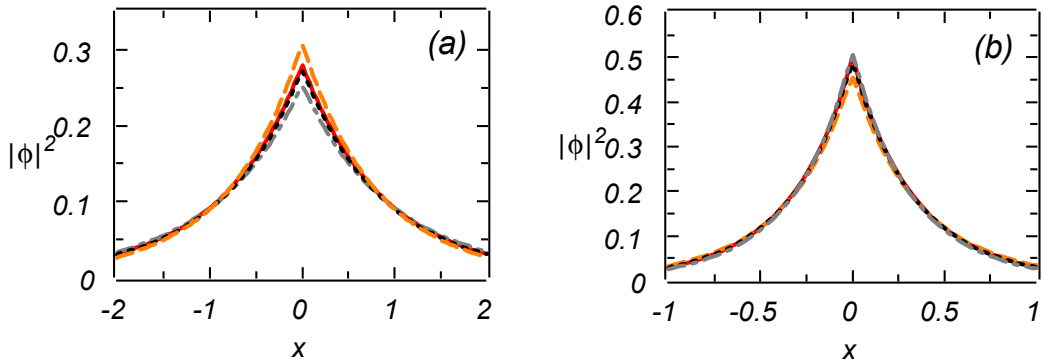}
\caption{Solid red lines display the exact solution (\protect\ref{sinh}) for
the mode pinned to the delta-functional pump. They are compared to a set of
numerically generated solutions produced with the help of the regularized
delta-function: dotted black lines pertain to mismatch parameter $\Delta $
taken exactly as per Eq. (\protect\ref{mism}); dashed orange lines pertain
to $\Delta \rightarrow 0.75\Delta $, and dashed-dotted gray lines pertain to
$\Delta \rightarrow 1.25\Delta $. The results displayed in panels (a) and
(b) are obtained for the self-focusing ($\protect\sigma =-1$) and
self-defocusing ($\protect\sigma =+1$) signs of the nonlinearity,
respectively. All these solutions are stable. Other parameters are $E_{0}=%
\protect\gamma =\Gamma =1$.}
\label{NE4}
\end{figure}

Because the above codimension-one analytical solution is valid only under
condition (\ref{mism}) imposed on the parameters, it is necessary to
investigate the \emph{structural stability} of the pinned modes against
departure from this condition. To this end, in Fig. \ref{NE4} we compare the
solutions (both analytical and numerically found ones) obtained with the
value of $\Delta $ selected as per Eq. (\ref{mism}), and their numerical
counterparts obtained with this $\Delta $ replaced by $0.75\Delta $ and $%
1.25\Delta $. We conclude that these considerable variations of $\Delta $
produce a weak effect on the solutions, i.e., they are structurally stable,
effectively representing generic pinned modes, rather than specially
selected ones.

While changes in the profiles of the solutions produced by the variation of $%
\Delta $ are relatively small, it is relevant to mention that the solutions
are more sensitive to the variation in the case of the self-focusing than in
the defocusing cas\textbf{e.}

Finally, systematic simulations of the perturbed solutions corroborate the
stability of all the numerical solutions emulating the analytically
predicted modes pinned to the delta-function. In fact, all the solutions are
strong \textit{attractors}, as direct simulations demonstrate that Eq. (\ref%
{New_LuLe-1D}) readily generates precisely these states, starting from the
zero input, $\phi (x,0)=0$. This numerical result is important, because the
stability of the analytically found solutions cannot be explored in an
analytical form.

\subsection*{The two-dimensional Lugiato-Lefever equation.}

The 2D version of the 1D LL equation (\ref{LuLe-1D}) is
\begin{equation}
i\left( \gamma +\frac{\partial }{\partial t}\right) \phi =\left[ -\frac{1}{2}%
\nabla _{\perp }^{2}+\Delta +\sigma |\phi |^{2}\right] \phi +E(x,y),
\label{LuLe}
\end{equation}%
where $\nabla _{\perp }^{2}=\frac{\partial ^{2}}{\partial x^{2}}+\frac{%
\partial ^{2}}{\partial y^{2}}$, and the cubic loss is not included here ($%
\Gamma =0$), as, unlike the exact 1D solutions, this term is not necessary
for finding 2D solutions reported here. Further, one may fix here $\gamma =1$
by means of rescaling. Below, we consider the Gaussian 2D shape of the pump,
given by
\begin{equation}
E(x,y)=\frac{P_{0}}{\sqrt{\pi }\eta }\exp \left( -\frac{x^{2}+y^{2}}{2\eta
^{2}}\right) ,  \label{pump}
\end{equation}%
where $P_{0}$ is the pump's integral intensity, and parameter $\eta $
controls its width.

\begin{figure}[tb]
\centering \includegraphics[width=1.\columnwidth]{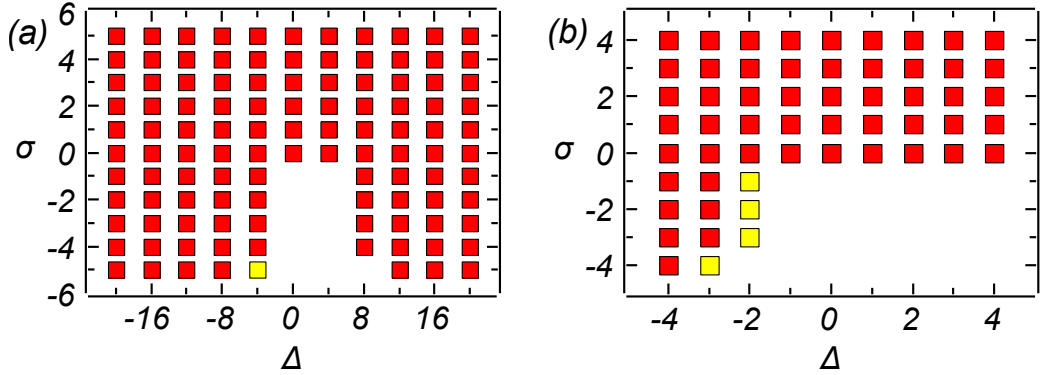}
\caption{Existence diagrams for stable 2D modes in the plane of parameters ($%
\Delta ,\protect\sigma $), as produced by direct simulations of Eq. (\protect
\ref{LuLe}), with pump's parameters $P_{0}=10$ and $\protect\eta =1$. Panel
(a) covers the range of $\protect\sigma \in \lbrack -5,5]$ and $\Delta \in
\lbrack -20,20]$, while (b) is a zoom of (a) for $\protect\sigma \in \lbrack
-4,4]$ and $\Delta \in \lbrack -4,4]$. The region covered by red boxes is
populated by single-peak (bell-shaped) modes (see Fig. \protect\ref{E4}),
while yellow boxes designate parameters at which the shape of the modes is
crater-shaped, featuring the maximum local power at a finite difference from
the center, see an example in Fig. \protect\ref{E6} below.}
\label{E1}
\end{figure}

In Fig. \ref{E1} we display the existence diagram of stable solutions
produced by direct simulations of Eq. (\ref{LuLe}) in the plane of the
mismatch and nonlinearity coefficients, ($\Delta ,\sigma $), for fixed
pump's parameters, $P_{0}=10$ and $\eta =1$. Light yellow boxes denote
values of parameters at which stable 2D\ solutions are crater-shaped (see
Fig. \ref{E6} below), while red boxes correspond to the single-peak
(bell-shaped) solutions, as shown below in Fig. \ref{E4}. As concerns
variational equations (\ref{eq1})-(\ref{eq3}) (see Methods), their
physically relevant solutions, corresponding to $B>0$, have been found, by
means of the relaxation method, for all values of parameters covering the
range of $\sigma \in \lbrack -5,5]$ and $\Delta \in \lbrack -20,20]$, the
respective picture essentially coinciding with one displayed on the basis of
the full numerical solution in Fig. \ref{E1}(a)

\begin{figure}[tb]
\centering \includegraphics[width=1.\columnwidth]{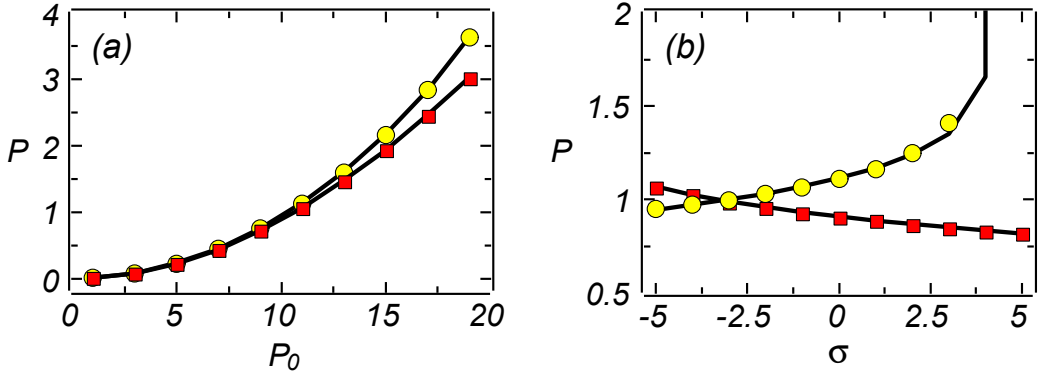}
\caption{(a) The integral power of the 2D modes, $P$, \textit{vs} the pump's
amplitude, $P_{0}$, for $\Delta =10$ and $\protect\sigma =-1$, shown by the
line with yellow circles, and $\protect\sigma =+1$, the line with red boxes.
(b) The integral power \textit{vs} $\protect\sigma $, pertaining to $P_{0}=10
$ and $\Delta =-10$ or $\Delta =+10$, shown by lines with yellow circles or
red boxes, respectively. Numerical and variational solutions are
indistinguishable in the range shown in the plots. Other parameters are $%
\protect\gamma =\protect\eta =1$.}
\label{E3}
\end{figure}

In Figs. \ref{E3}(a) and \ref{E3}(b) we display the integral power of the
confined 2D modes, $P$, defined as per Eq. (\ref{Norm}), as a function of
the pump's amplitude $P_{0}$ and nonlinearity strength $\sigma $,
respectively (see Eq. (\ref{pump})). Note that in Fig. \ref{E3}(a) the
analytical results, produced by the variational \textit{ansatz} (\ref{ansatz}%
), and their numerical counterpart, obtained from direct simulations of Eq. (%
\ref{LuLe}), are very close to each other. We observe that, in the
self-focusing case ($\sigma =-1$, shown by the line with circles), the
integral power is slightly larger than in the self-defocusing case ($\sigma
=1$, shown by the line with boxes). In Fig. \ref{E3}(b), the abrupt growth
of the power for $\Delta =-10$ at $\sigma >4$ make the numerical solutions
unstable.

\begin{figure}[tb]
\centering \includegraphics[width=1.\columnwidth]{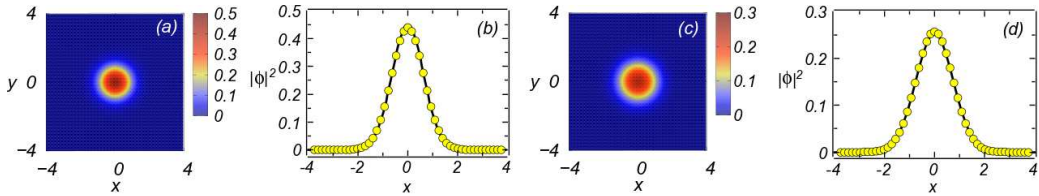}
\caption{Profiles of stable 2D modes, $|\protect\phi \left( x,y\right) |^{2}
$, as produced by direct simulations of Eq. ( \protect\ref{LuLe}) (at $t=100$%
), for (a) $\Delta =-10$ and (c) $\Delta =10$. Displayed in panels (b) and
(d) are transverse profiles, $|\protect\phi (x,0)|^{2}$, corresponding to
the 2D shapes shown in (a) and (c), respectively. Lines in (b) and (d)
depict the approximate analytical solution based on \textit{ansatz} (\protect
\ref{ansatz}), while chains of yellow circles represent the numerical
solution. Other parameters are $P_{0}=10$ and $\protect\sigma =\protect%
\gamma =\protect\eta =1$.}
\label{E4}
\end{figure}

Generic examples of the local-power profiles, $|\phi |^{2}$, for the 2D
modes, obtained from direct simulations (at $t=100$) for two different
values of $\Delta $, and the comparisons with the corresponding approximate
analytical solutions, based on \textit{ansatz} (\ref{ansatz}), are displayed
in Fig. \ref{E4}. Actually, the numerical and analytical profiles are
indistinguishable at these values of $\Delta $, in accordance with the above
results which also demonstrated very good agreement of the analytical
predictions with the numerical counterparts at large values of $\Delta $.
However, at small values of $\Delta $, the numerical solutions feature a
strong increase in the norm and may become unstable. In this case, the
analytical approximation is not relevant.

\begin{figure}[tb]
\centering \includegraphics[width=1.\columnwidth]{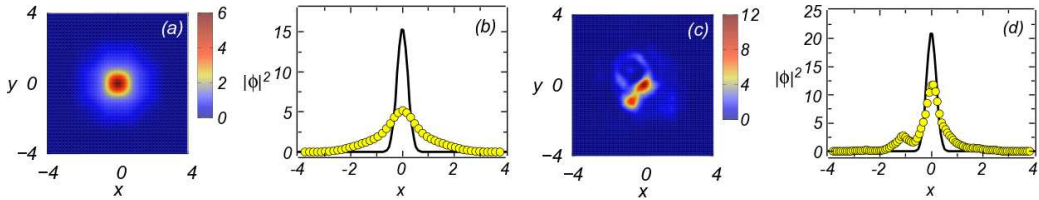}
\caption{Profiles of 2D solutions, $|\protect\phi \left( x,y\right) |^{2}$,
and the corresponding transverse profile, $|\protect\phi (x,0)|^{2}$, for $%
\Delta =-1$ in (a)-(b), and $\Delta =+1$ in (c)-(d). In (b) and (d), solid
black lines represents the analytical results, that were used as inputs for
the direct simulations. Results of the simulations (at $t=10$) are shown by
yellow circles. Other parameters are $\protect\sigma =-1$, $P_{0}=10$, and $%
\protect\gamma =\protect\eta =1$.}
\label{E5}
\end{figure}

The situation in a parameter region where stable stationary modes are absent
(see Fig. \ref{E1}) is illustrated by numerically generated solutions (at $%
t=10$) displayed in Figs. \ref{E5}(a,b) for $\sigma =-1$ and $\Delta =-1$,
and in Figs. \ref{E5}(c)-(d) for $\sigma =-1$ and $\Delta =+1$. Due to the
instability of the numerical solutions, the analytical predictions are not
relevant in this case.

\begin{figure}[tb]
\centering \includegraphics[width=1.\columnwidth]{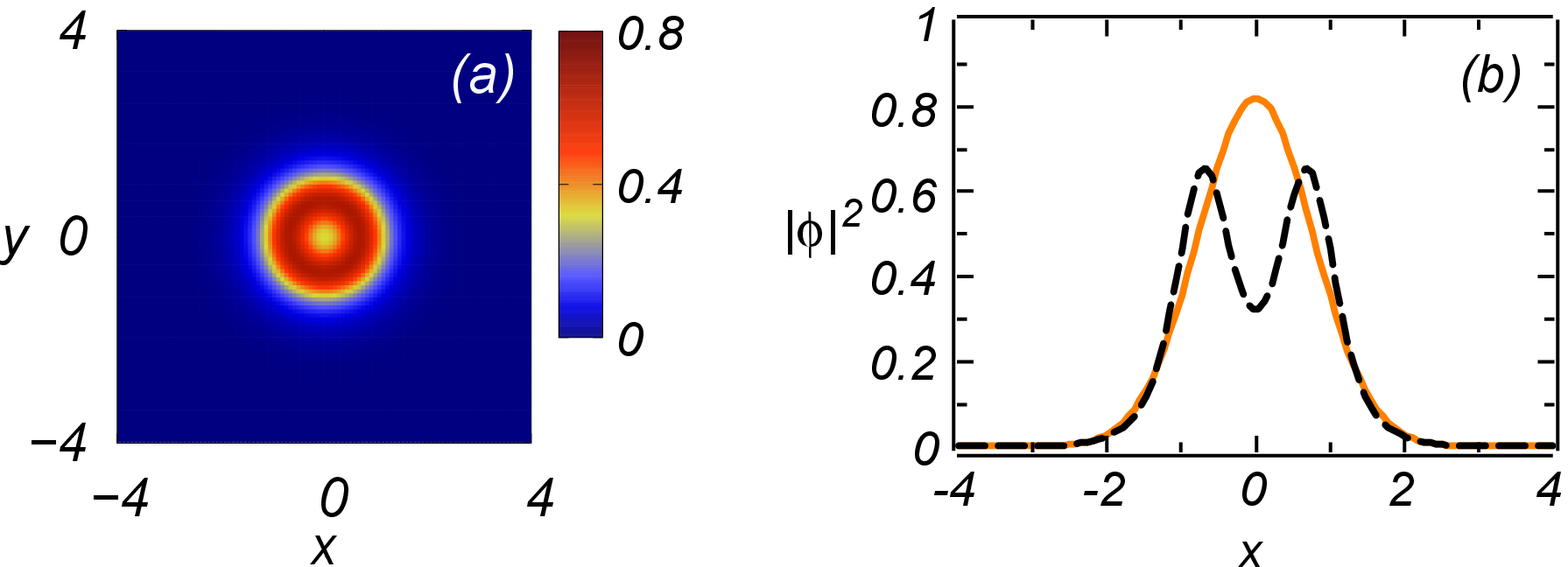}
\caption{(a) The same as in Fig. \protect\ref{E4}(a), but with $\protect%
\sigma =-5$ and $\Delta =-4$. (b) The black dashed line is the static
profile, $\left\vert \protect\phi \left( x,0\right) \right\vert ^{2}$, of
the numerically generated crater-shaped mode (at $t=100$). The orange solid
line shows a formal prediction of the variational approximation for these
values of the parameters.}
\label{E6}
\end{figure}

As mentioned above, in a small part of their existence region (covered by
yellow squared in Fig. \ref{E1}), numerically found stable 2D modes feature
a crater-like shape, with the maximum of the local power attained at a
finite difference from the center, see an example in Fig. \ref{E6}.
Obviously, the analytical approximation based on \textit{ansatz} (\ref%
{ansatz}) cannot reproduce this shape.

\section*{Discussion}

\qquad The modifications of the well-known 1D and 2D LL (Lugiato-Lefever)
equation introduced in this work, with tightly localized pump, make it
possible to create new stable confined modes, which are of interest in terms
of the use of the LL equations as models of the pattern formation in
nonlinear dissipative media. They may also be used to design compact pixels
that can be created in cavities modeled by the LL equations. The present
work is based on the combination of analytical and numerical methods, in the
1D and 2D geometries alike, the analytical parts helping to achieve a deeper
insight into the variety of steady-state confined modes produced by the LL
equations.

In the 1D geometry, we have first developed a simple perturbation theory,
based on the usual \textrm{sech} \textit{ansatz} (\ref{soliton}), in the
case of weak pump and loss. Other results have produced a family of exact
analytical solutions, assuming that the tightly focused gain is represented
by the delta-function, while the self-interaction may have both focusing and
defocusing signs ($\sigma <0$ and $\sigma >0$, respectively). The analytical
form of the solution is given by Eqs. (\ref{sinh})-(\ref{lambda^2}), under
the condition that the mismatch, $\Delta $, takes the specially selected
value (\ref{mism}), and the cubic nonlinear term, which represents
two-photon losses in the optical medium (with rate $\Gamma $), is present.
Furthermore, numerical results, displayed in Fig. \ref{NE4}, corroborate the
structural stability of the codimension-one analytical solutions, because
the deviation of $\Delta $ from the spacial value (\ref{mism}) leads to weak
variation of the stable pinned solutions. Most essential parameters which
control the shape of the 1D pinned modes are two nonlinearity coefficient, $%
\sigma $ and $\Gamma $. Characteristic features of the solution is the cusp
at the center, and the phase structure (chirp). A remarkable fact is that
the exact solutions are very close to their numerical counterparts, produced
by the localized pump shaped as a regularized delta-function, and the family
of the so generated 1D modes is entirely stable. In fact, the proximity of
the numerical and analytical solutions additionally confirms the structural
stability of the latter. A noteworthy (and counter-intuitive) feature of the
1D modes is that they \textit{shrink} with the increase of the strength of
the self-defocusing nonlinearity. The 1D solution produced by the analysis
may help to find similar states in more general pattern-formation models.

In the 2D geometry with the pump applied at a small Gaussian-shape
\textquotedblleft hot spot", systematic numerical results are reported in
the combination with approximate analytical findings produced by the
variational approximation. A vast stability area in the system's parameter
space has been found, the most essential parameters being the
above-mentioned mismatch and nonlinearity coefficients, $\Delta $ and $%
\sigma $ (the 2D system is considered without the two-photon loss, $\Gamma
=0 $, as its presence is not a necessary condition for finding the relevant
solutions). In most cases, the 2D modes pinned to the \textquotedblleft hot
spot" feature a single-peak (bell-shaped) structure, which is stable, and is
well approximated by the variational \textit{ansatz}. In a small part of the
parameter space, 2D stable modes feature a crater-like shape, with the
maximum local power found at a finite distance from the center. In another
small part of the parameter space, 2D modes are unstable.

As an extension of the analysis, it may be interesting to use numerical
methods to construct modes pinned to a set of two mutually symmetric 1D or
2D hot spots, cf. a similar configuration elaborated for the 1D CGL equation
in Ref. \cite{Kwok}. In particular, in the case of the self-focusing
nonlinearity, $\sigma<0$, one may expect spontaneous symmetry breaking
between peaks attached to the two pump maxima. On the other hand, in the 2D
geometry it may also be interesting to introduce ring-shaped pump, which may
give rise to confined modes with a vortex structure, cf. a similar
consideration for the 2D CGL equation in Ref. \cite{Huang_OL13}. A
possibility of spontaneous breaking of the axial symmetry in vortex modes
may be addressed too, following the pattern of the analysis performed in the
framework of the CGL equation \cite{stars}.

\section*{Methods}

\qquad{}\textbf{The variational approach.} Firstly, we define $%
\phi(x,y,t)\equiv\Phi(x,y,t)\exp\left(-\gamma t\right)$, casting Eq. (\ref%
{LuLe}) in the form of
\begin{equation}
i\frac{\partial}{\partial t}\Phi=\left[-\frac{1}{2}\nabla_{\perp}^{2}+%
\Delta+\sigma e^{-2\gamma t}|\Phi|^{2}\right]\Phi+Ee^{\gamma t}\;,
\label{PhiLL}
\end{equation}
which can be derived from a real time-dependent Lagrangian,
\begin{eqnarray}
L & = & \int\int dxdy\left\{ \frac{i}{2}\left(\Phi_{t}^{\ast}\Phi-\Phi^{%
\ast}\Phi_{t}\right)+\frac{1}{2}\left(|\Phi_{x}|^{2}+|\Phi_{y}|^{2}\right)%
\right.  \notag \\
& + & \left.\Delta|\Phi|^{2}+\frac{\sigma}{2}e^{-2\gamma
t}\,|\Phi|^{4}+Ee^{\gamma t}\left(\Phi^{\ast}+\Phi\right)\right\} \,.
\label{LLL}
\end{eqnarray}
Note that the following exact power-balance equation is produced by Eq. (\ref%
{LuLe}):
\begin{equation}
\frac{dP}{dt}=-2\gamma P-2\iint\mathrm{Im}\{\phi(x,y,t)\}E(x,y)dxdy,
\label{PB}
\end{equation}
for the integral power defined as
\begin{equation}
P=\int\int|\phi(x,y,t)|^{2}dxdy,  \label{Norm}
\end{equation}
cf. the 1D counterpart given by Eq. (\ref{balance}).

For the variational approximation, we use the 2D isotropic Gaussian \textit{%
ansatz }\cite{Anderson},
\begin{equation}
\Phi=e^{\gamma t}A(t)\exp\left[-\left(B(t)-iC(t)\right)(x^{2}+y^{2})\right],
\label{ansatz}
\end{equation}
where $A$, $B$ and $C$ are real variational parameters, subject to obvious
constraint $B>0$. Next, substituting the \textit{ansatz} in Eq. (\ref{LLL})
and performing the integration, we arrive at the following effective
Lagrangian:
\begin{eqnarray}
L_{\mathrm{eff}} & = & \frac{\pi}{2}e^{2\gamma t}\left\{ \frac{A^{2}}{2B^{2}}%
\frac{dC}{dt}+\left[1+\frac{\Delta}{B}+\frac{C^{2}}{B^{2}}\right]A^{2}\right.
\notag \\
& + & \left.\frac{\sigma A^{4}}{4B}+\frac{8\eta\mathit{P_{0}}%
\left(1+2B\eta^{2}\right)A}{1+\left(4B^{2}+4C^{2}\right)\eta^{4}+4B\eta^{2}}%
\right\} .  \label{Leff}
\end{eqnarray}
The variational (Euler-Lagrange) equations following from Lagrangian (\ref%
{Leff}), $\partial L_{\mathrm{eff}}/\partial\left(A,B,C\right)=0$, are
\begin{equation}
\sigma A^{3}+\frac{\left[4B(B+\Delta)+4C^{2}\right]}{2B}A+\frac{8\eta\mathit{%
P_{0}}B^{2}\left(1+2B\eta^{2}\right)}{\sqrt{\pi}\left[4B^{3}\eta^{4}+4B^{2}%
\eta^{2}+\left(4\eta^{4}C^{2}+1\right)B\right]}=0,  \label{eq1}
\end{equation}
\begin{equation}
\sigma BA^{3}+4\left(\Delta B+2C^{2}\right)A+\frac{8\eta^{3}\mathit{P_{0}}%
B^{3}\left[1+4\left(B^{2}-C^{2}\right)\eta^{4}+4B\eta^{2}\right]}{\sqrt{\pi}%
\left(1+4\left(B^{2}+C^{2}\right)\eta^{4}+4B\eta^{2}\right)^{2}}=0,
\label{eq2}
\end{equation}
\begin{equation}
\frac{\pi\left(2C-\gamma\right)A}{2B^{2}}-\frac{32\eta^{5}\mathit{P_{0}}%
\left(1+2B\eta^{2}\right)C}{\sqrt{\pi}\left[1+4\left(B^{2}+C^{2}\right)%
\eta^{4}+4B\eta^{2}\right]^{2}}=0.  \label{eq3}
\end{equation}

\textbf{Numerical simulations.} To solve the one-dimensional equation (\ref%
{New_LuLe-1D}) numerically, we made use of the regularized delta-function
based on the usual Gaussian expression (see Ref. \cite{Malomed_JOSAB14} and
references therein):
\begin{equation}
\widetilde{\delta}(x)=(\sqrt{\pi}w)^{-1}\exp(-x^{2}/w^{2}),  \label{delta}
\end{equation}
with finite width $w$.

We employed a fourth-order split-step method to solve Eqs. (\ref{New_LuLe-1D}%
) and (\ref{LuLe}), starting from the zero input, $\phi (x,0)=0$. An output
was categorized as a stable mode if it maintained a static profile for a
long time ($t\sim 1000$, which corresponds $\gtrsim 100$ characteristic
diffraction times). In most simulations, the spatial and temporal steps were
fixed as $\Delta x=0.04$ and $\Delta t=0.001$. To produce the numerical
results for the 1D LL equation, shown in Fig. \ref{1DF2}, we chose values of
$w$ such that the resultant integral power was different from the analytical
counterpart, corresponding to exact solution (\ref{sinh}), by no more than $%
3\%$.

To solve Eqs. (\ref{eq1})-(\ref{eq3}) for $A$, $B$, and $C$, produced by the
variational approximation, we used a relaxation method with a fixed error
constraint of $10^{-6}$. Then, the so found values were inserted in \textit{%
ansatz} (\ref{ansatz}) to produce the full variational approximation for the
2D modes. Lastly, the above-mentioned scaled value of the dissipation
parameter, $\gamma =1$, was set in all the simulations.

\section*{Acknowledgements}

W.B.C. acknowledges financial support from the Brazilian agencies CNPq
(grant \#458889/2014-8) and the National Institute of Science and Technology
(INCT) for Quantum Information. The work of B.A.M. is supported, in part, by
grant No. 2015616 from the joint program in physics between NSF (US) and
Binational (US-Israel) Science Foundation, and by grant No. 1287/17 from the
Israel Science Foundation. The work of L.S. is supported, in part, by BIRD
project No. 164754 of University of Padova.

\section*{Author contributions}

The numerical part has been carried out by W.B.C. Analytical considerations
were chiefly performed by B.A.M. and L.S. All the authors have contributed
to drafting the manuscript.

\section*{Additional information}

\textbf{Competing interests}: The authors declare no competing financial
interests.

\end{document}